# The Perception of Time, Risk And Return During Periods Of Speculation

Emanuel Derman
Goldman, Sachs & Co.
January 10, 2002

*The goal of trading ... was to dart in and out of the electronic marketplace, making a series of small profits. Buy at 50 sell at 50 1/8. Buy at 50 1/8, sell at 50 1/4. And so on.*

*"My time frame in trading can be anything from ten seconds to half a day. Usually, it's in the five-to-twenty-five minute range."*

*By early 1999 ... day trading accounted for about 15% of the total trading volume on the Nasdaq.*

John Cassidy on day-traders, in "Striking it Rich."
The New Yorker, Jan. 14, 2002.

# Summary

What return should you expect when you take on a given amount of risk? How should that return depend upon other people's behavior? What principles can you use to answer these questions? In this paper, we approach these topics by exploring the consequences of two simple hypotheses about risk.

The first is a common-sense invariance principle: assets with the same perceived risk must have the same expected return. It leads directly to the well-known Sharpe ratio and the classic risk-return relationships of Arbitrage Pricing Theory and the Capital Asset Pricing Model.

The second hypothesis concerns the perception of time. We conjecture that in times of speculative excitement, short-term investors may instinctively imagine stock prices to be evolving in a time measure different from that of calendar time. They may perceive and experience the risk and return of a stock in *intrinsic time*, a dimensionless time scale that counts the number of trading opportunities that occur, but pays no attention to the calendar time that passes between them.

Applying the first hypothesis in the intrinsic time measure suggested by the second, we derive an alternative set of relationships between risk and return. Its most noteworthy feature is that, in the short-term, a stock's trading frequency affects its expected return. We show that short-term stock speculators will expect returns proportional to the *temperature* of a stock, where temperature is defined as the product of the stock's traditional volatility and the square root of its trading frequency. Furthermore, we derive a modified version of the Capital Asset Pricing Model in which a stock's excess return relative to the market is proportional to its traditional beta multiplied by the square root of its trading frequency.

We hope that this model will have some relevance to the behavior of investors expecting inordinate returns in highly speculative markets.

# Table of Contents

# 1 Overview

What should you pay for a given amount of risk? How should that price depend upon other people's behavior and sentiments? What principles can you use to help answer these questions?

These are old questions which led to the classic mean-variance formulation of the principles of modern finance[1], but have still not received a definitive answer. The original theory of stock options valuation[2] and its manifold extensions has been so widely embraced because it provides an unequivocal and almost sentiment-free prescription for the replacement of an apparently risky, unpriced asset by a mixture of other assets with known prices. But this elegant case is the exception. Most risky assets cannot be replicated, even in theory.

In this note I want to explore the consequences of two hypotheses. The first is a simple invariance principle relating risk to return: assets with the same perceived risk must have the same expected return. When applied to the valuation of risky stocks, it leads to results similar to those of the Capital Asset Pricing Model[3] and Arbitrage Pricing Theory[4]. Although the derivation here may not be the usual one, it provides a useful framework for further generalization.

The second hypothesis is a conjecture about an alternative way in which investors perceive the passage of time and the risks it brings. Perhaps, at certain times, particularly during periods of excited speculation, some market participants may, instinctively or consciously, pay significant attention to the rate at which trading opportunities pass, that is, to the stock's *trading frequency*. In excitable markets, the trading frequency may temporarily seem more important than the rate at which ordinary calendar time flows by.

The trading frequency of a stock implicitly determines an *intrinsic time* scale[5], a time whose units are ticked off by an imaginary clock that measures the passing of trading opportunities for that particular stock. Each stock has its own relationship between its intrinsic time and calendar time, determined by its trading frequency. Though trading frequencies vary with time in both systematic and random ways, in this paper we will use only the average trading frequency of the stock, and ignore any contributions from its fluctuations.

The combination of these two hypotheses – that similar risks demand similar returns, and that short-term investors look at risk and return in terms of intrinsic time – leads to alternative relationships between risk and return. In the short run, expected return is proportional to the *temperature* of the stock, where temperature is the product of the standard volatility and the square root of trading frequency. Stocks that trade more frequently produce a short-term expectation of greater returns. (This can only be true in the short run. In the long run, the ultimate return generated by a company will depend on its profitability and not on its trading frequency.) We will derive and elaborate on these results in the main part of this paper.

---

1. [Markowitz 1952].
2. [Black 1973] and [Merton 1973].
3. See Chapter 7 of [Luenberger 1999] for a summary of the Sharpe-Lintner-Mossin Capital Asset Pricing Model.
4. [Ross 1976].
5. See for example [Clark 1973] and [Müller 1995], who used intrinsic time to mean the measure that counts as equal the elapsed time between trades.

My motivation for these re-derivations and extensions is three-fold. First, I became curious about the extent to which interesting and relevant macroscopic results about financial risk and reward could be derived from a few basic principles. Here I was motivated by nineteenth-century thermodynamics, where many powerful and practical constraints on the production of mechanical energy from heat follow from a few easily stated laws; also by special relativity, which is not a physical theory but rather a meta-principle about the form of all possible physical theories. In physics, a foundation of macroscopic understanding has traditionally preceded microscopic modeling, Perhaps one can find analogous principles on which to base microscopic finance.

Second, I became interested in the notion that the observed lack of normality in the distribution of calendar-time stock returns might find some of its origins in the randomly varying time between the successive trades of a stock[6]. Some authors have suggested that the distribution of a stock's returns, as measured per unit of intrinsic time, may more closely resemble a normal distribution.

Finally, in view of the remarkable returns of technology and internet stocks over the past few years, I had hoped to find some new (perhaps behavioral) relationships between risk and reward that might apply to these high-excitement markets. Traditional approaches have sought to regard these temporarily high returns as either the manifestation of an irrational greed on the part of speculators, or else as evidence of a concealed but justifiable optionality in future payoffs[7]. Since technology markets in recent years have been characterized by periods of rapid day-trading, perhaps intrinsic time, in taking account of the perception of the rate at which trading opportunities present themselves, is a parameter relevant to sentiment and valuation.

This paper proceeds as follows. In Section 2 I formulate the first hypothesis, the invariance principle for valuing stocks, and then apply it to four progressively more realistic and complex cases. These are (i) uncorrelated stocks with no opportunity for diversification, (ii) uncorrelated stocks which can be diversified, (iii) stocks which are correlated with the overall market but provide no opportunity for diversification, and finally, (iv) diversifiable stocks which are correlated with a single market factor. In this final case, the invariance principle leads to the traditional Capital Asset Pricing Model.

In Section 3 I reformulate the invariance principle in intrinsic time. The main consequence is that a stock's trading frequency affects its expected return. Short-term stock speculators will expect the returns of stocks uncorrelated with the market to be proportional to their *temperature*. "Hotter" stocks have higher expected returns. For stocks correlated with the overall market, a frequency-adjusted Capital Asset Pricing Model holds, in which a stock's excess return relative to the market is proportional to its traditional beta multiplied by the square root of its trading frequency.

---

6. For examples, see [Clark 1973], [Geman 1996], [Madan 1998] and [Plerou 2000].
7. See [Schwartz 2000] and [Posner 2000] for examples of the hidden-optionality models of internet stocks.

In Section 4, I briefly examine how this theory of intrinsic time can be extended to options valuation, and can thereby perhaps account for some part of the volatility skew.

The Appendix provides a simple illustration of how market bubbles can be caused by investors who, while expecting the returns traditionally associated with observed volatility, instead witness the returns induced by short-term temperature-sensitive speculators.

I hope that the macroscopic models described below may provide a description of the behavior of stock prices during so-called market bubbles.

# 2 A Simple Invariance Principle and its Consequences

## 2.1 A Stock's Risk and Return

Suppose the market consists of (i) a single risk-free bond B of price $B$ that provides a continuous riskless return $r$, and (ii) the stocks of $N$ different companies, where each company $i$ has issued $n_i$ stocks of current market value $S_i$. Here, and in what follows, we use capital letters like B and $S_i$ to denote the names of securities, and the italicized capitals $B$ and $S_i$ to denote their prices.

I assume (for now) that a stock's only relevant information-bearing parameter is its riskiness, or rather, its *perceived* riskiness[8]. Following the classic approach of Markowitz, I assume that the appropriate measures of stock risk are volatility and correlation. Suppose that all investors assume that each stock price will evolve lognormally during the next instant of time $dt$ in the familiar continuous way, so that

$$\frac{dS_i}{S_i} = \mu_i dt + \sigma_i dZ_i \qquad [2.1]$$

Here $\mu_i$ represents the value of the expected instantaneous return (per unit of calendar time) of stock $S_i$, and $\sigma_i$ represents its volatility. We use $\rho_{i,j}$ to represent the correlation between the returns of stock $i$ and stock $j$. The Wiener processes $dZ_i$ satisfy

$$dZ_i^2 = dt$$
$$dZ_i dZ_j = \rho_{ij} dt \qquad [2.2]$$

We have assumed that stocks undergo the traditional lognormal model of evolution. To some extent this assumption is merely a convenience. If you believe in a more complex evolution of stock prices, there is a correspondingly more complex version of many of the results derived below.

## 2.2 The Invariance Principle

I can think of only one virtually inarguable principle that relates the expected returns of different stocks, namely that

> *Two portfolios with the same perceived risk should have the same expected return.*

In the next subsection we will explore the consequences of this principle, assuming that both return and risk are evaluated conventionally, in calendar time. In later sections, we will also examine the possibility that what matters to investors is not risk and return in calendar time, but rather, risk and return as measured in *intrinsic time.*

---

8. I say "for now" in this sentence because in Section 3 I will loosen this assumption by also allowing the expected time between trading opportunities to carry information.

We will identify the word "risk" with volatility, that is, with the annualized standard deviation of returns. However, even if risk were measured in a more complex or multivariate way, we would still assume the above invariance principle to be valid, albeit with a richer definition of risk.

This invariance principle is a more general variant of the *law of one price* or the principle of *no riskless arbitrage,* which dictates, more narrowly, that only two portfolios with exactly the same future payoffs in all states of the world should have the same current price. This latter principle is the basis of the theory of derivatives valuation.

Our aim from now on will be to exploit this simple principle – that stocks with the same perceived risk must provide the same expected return – in order to extract a relationship between the prices of different stocks. We begin by applying the principle in a market (or market sector) with a small number of uncorrelated stocks where no diversification is available.

## 2.3 Uncorrelated Stocks in an Undiversifiable Market

Consider two stocks S and P whose prices are assumed to evolve according to the stochastic differential equations

$$\frac{dS}{S} = \mu_S dt + \sigma_S dZ_S$$

$$\frac{dP}{P} = \mu_P dt + \sigma_P dZ_P$$

[2.3]

Here $\mu_i$ is the expected value for the return of stock $i$ in calendar time and $\sigma_i$ is the return volatility. For convenience we assume that $\sigma_P$ is greater than $\sigma_S$. If calendar time is measured in years, then the units of $\mu$ are percent per year and the units of $\sigma$ are percent per square root of a year. The dimension of $\mu$ is [time]$^{-1}$ and that of $\sigma$ is [time]$^{-1/2}$.

The riskless bond B is assumed to compound annually at a rate $r$, so that

$$\frac{dB}{B} = r dt$$

[2.4]

An investor faced with buying stock S or stock P needs to be able to decide between the attractiveness of earning (or, more accurately, expecting to earn) $\mu_S$ with risk $\sigma_S$ vs. earning $\mu_P$ with risk $\sigma_P$. Which of these alternatives provides a better deal?

To answer this, we note that, at any time, by adding some investment in a riskless (zero-volatility) bond B to the riskier stock P (with volatility $\sigma_P$), we can create a portfolio of lower volatility. More specifically, one can instantaneously construct a portfolio V consisting of $w$ shares of P and $1 - w$ shares of B, with $w$ chosen so that the instantaneous volatility of V is the same as the volatility of S.

We write

$$V = wP + (1-w)B \qquad [2.5]$$

Then, from Equations 2.3 and 2.4,

$$\frac{dV}{V} = \mu_V(t)dt + \sigma_V(t)dZ_P \qquad [2.6]$$

where

$$\mu_V = \frac{w\mu_P P + (1-w)rB}{wP + (1-w)B}$$

$$\sigma_V = \frac{wP\sigma_P}{wP + (1-w)B} \qquad [2.7]$$

are the expected return and volatility of V, conditioned on the values of P and B at time $t$.

We now choose $w$ such that V and S have the same instantaneous volatility $\sigma_S$. Equating $\sigma_V$ in Equation 2.7 to $\sigma_S$ we find that $w$ must satisfy

$$w = \frac{\sigma_S B}{\sigma_S B + (\sigma_P - \sigma_S)P} \qquad [2.8]$$

where the dependence of the prices $P$ and $B$ on the time parameter $t$ is suppressed for brevity. It is convenient to write the equivalent expression

$$\frac{1}{w} = 1 + \frac{P}{B}\left(\frac{\sigma_P}{\sigma_S} - 1\right) \qquad [2.9]$$

Since V and S carry the same instantaneous risk, our invariance principle demands that they provide the same expected return, so that $\mu_V = \mu_S$. Equating $\mu_V$ in Equation 2.7 to $\mu_S$ we find that $w$ must also satisfy

$$w = \frac{(\mu_S - r)B}{(\mu_S - r)B + (\mu_P - \mu_S)P} \qquad [2.10]$$

or, equivalently,

$$\frac{1}{w} = 1 + \frac{P}{B}\frac{(\mu_P - \mu_S)}{(\mu_S - r)} \qquad [2.11]$$

where the explicit time-dependence is again suppressed.

By equating the right-hand sides of Equations 2.9 and 2.11, and separating the S- and P-dependent variables, one can show that

$$\frac{\mu_S - r}{\sigma_S} = \frac{\mu_P - r}{\sigma_P} \qquad [2.12]$$

Since the left-hand side of Equation 2.12 depends only on stock S and the right-hand side depends only on stock P, they must each be equal to a stock-independent constant $\lambda$. Therefore, for any portfolio $i$,

$$\frac{\mu_i - r}{\sigma_i} = \lambda \qquad [2.13]$$

or

$$\mu_i - r = \lambda \sigma_i \qquad [2.14]$$

Equation 2.14 dictates that the excess return per unit of volatility, the well-known Sharpe ratio $\lambda$, is the same for all stocks. Nothing yet tells us the value of $\lambda$. Perhaps a more microscopic model[9] of risk and return can provide a means for calculating $\lambda$. The dimension of $\lambda$ is [time]$^{-1/2}$, and so a microscopic model of this kind must contain at least one other parameter with the dimension of time[10].

## 2.4 Uncorrelated Stocks in a Diversifiable Market

An investor who can own only an individual stock $S_i$ is exposed to its price risk. But, if large numbers of stocks are available, diversification can decrease the risk. Suppose that at some instant the investor buys a portfolio V consisting of $l_i$ shares of each of $L$ different stocks, so that the portfolio value $V$ is given by

$$V = \sum_{i=1}^{L} l_i S_i \qquad [2.15]$$

Then the evolution of the value of this portfolio satisfies

$$dV = \sum_{i=1}^{L} l_i dS_i = \sum_{i=1}^{L} l_i S_i (\mu_i dt + \sigma_i dZ_i) = \left( \sum_{i=1}^{L} l_i S_i \mu_i \right) dt + \sum_{i=1}^{L} l_i S_i \sigma_i dZ_i$$

The instantaneous return on the portfolio is

$$\frac{dV}{V} = \left(\sum_{i=1}^{L} w_i \mu_i\right) dt + \sum_{i=1}^{L} w_i \sigma_i dZ_i \qquad [2.16]$$

where

$$w_i = (l_i S_i) \Big/ \left(\sum_{i=1}^{L} l_i S_i\right) \qquad [2.17]$$

is the initial capitalization weight of stock $i$ in the portfolio V, and

$$\sum_{i=1}^{L} w_i = 1 .$$

---

9. What I have in mind is the way in which measured physical constants become theoretically calculable in more fundamental theories. An example is the Rydberg constant that determines the density of atomic spectral lines, which, once Bohr developed his theory of atomic structure, was found to be a function of the Planck constant, the electron charge and its mass.

10. Here is a brief look ahead: one parameter whose dimension is related to time is trading frequency. In Section 3 we develop an alternative model in which the Sharpe ratio $\lambda$ is found to be proportional to the square root of the trading frequency.

According to Equation 2.16, the expected return of portfolio V is

$$\mu_V = \sum_{i=1}^{L} w_i \mu_i \qquad [2.18]$$

and the variance per unit time of the return on the portfolio is given by

$$\sigma_V^2 = \sum_{i,j=1}^{L} w_i w_j \rho_{ij} \sigma_i \sigma_j \qquad [2.19]$$

You can rewrite Equation 2.19 as

$$\sigma_V^2 = \sum_{i=1}^{L} w_i^2 \sigma_i^2 + \sum_{i \neq j} w_i w_j \rho_{ij} \sigma_i \sigma_j$$

The first sum consists of $L$ terms, the second of $L(L-1)$ terms. If all the stocks in V are approximately equally weighted so that $w_i \sim O(1/L)$, and if, on average, their returns are uncorrelated with each other, so that $\rho_{ij} < O(1/L)$, then

$$\sigma_V^2 \sim O(1/L) \to 0 \ \text{ as } \ L \to \infty \qquad [2.20]$$

So, by combining an individual stock with large numbers of other uncorrelated stocks, you can create a portfolio whose asymptotic variance is zero. In this limit, V is riskless. If the invariance principle holds not only for individual stocks but also for all portfolios, then applying Equation 2.14 to the portfolio V in this limit leads to

$$\mu_V - r \sim \lambda \sigma_V \sim 0 \qquad [2.21]$$

By substituting Equation 2.18 into Equation 2.21 we obtain

$$\sum_{i=1}^{L} w_i(\mu_i - r) \sim 0$$

We now use Equation 2.14 for each stock to replace $(\mu_i - r)$ by $\lambda\sigma_i$ in the above equation, and so obtain

$$\lambda\left[\sum_{i=1}^{L} w_i\sigma_i\right] \sim 0$$

To satisfy this demands that

$$\lambda \sim 0 \qquad [2.22]$$

Setting $\lambda \sim 0$ in Equation 2.13 implies that

$$\mu_i \sim r \qquad [2.23]$$

Therefore, in a diversifiable market, all stocks, irrespective of their volatility, have an expected return equal to the riskless rate, because their risk can be eliminated by incorporating them into a large portfolio. Equation 2.23 is a simplified version of the Capital Asset Pricing Model in a hypothetical world in which there is no market factor and all stocks are, on average, uncorrelated with each other.

## 2.5 Undiversifiable Stocks Correlated with One Market Factor

In the previous section we dealt with stocks whose average joint correlation was zero. Now we consider a situation that more closely resembles the real world in which all stocks are correlated with the overall market.

Suppose the market consists of N companies, with each company $i$ having issued $n_i$ stocks of current market value $S_i$. Suppose further that there is a traded index M that represents the entire market. Assume that the price of M evolves lognormally according to the standard Wiener process

$$\frac{dM}{M} = \mu_M dt + \sigma_M dZ_M \qquad [2.24]$$

where $\mu_M$ is the expected return of M and $\sigma_M$ is its volatility. We still assume that the prices of any stock $S_i$ and the price of the riskless bond B evolve according to the equations

$$\frac{dS_i}{S_i} = \mu_i dt + \sigma_i dZ_i$$

$$\frac{dB}{B} = r dt$$

$$[2.25]$$

where

$$dZ_i = \rho_{iM} dZ_M + \sqrt{1 - \rho_{iM}^2}\, \varepsilon_i \qquad [2.26]$$

Here $\varepsilon_i$ is a random normal variable that represents the residual risk of stock $i$, uncorrelated with $dZ_M$. We assume that both $\varepsilon_i^2 = dt$ and $dZ_M^2 = dt$, so that $dZ_i^2 = dt$ and $dZ_i dZ_M = \rho_{iM} dt$.

Because all stocks are correlated with the market index M, you can create a market-neutral version of each stock $S_i$ by shorting just enough of M to remove all market risk. Let $\tilde{S}_i$ denote the value of the market-neutral portfolio corresponding to the stock $S_i$, namely

$$\tilde{S}_i = S_i - \Delta_i M \qquad [2.27]$$

From Equations 2.24 − 2.27, the evolution of $\tilde{S}_i$ is given by

$$
\begin{aligned}
d\tilde{S}_i &= dS_i - \Delta_i dM \\
&= S_i(\mu_i dt + \sigma_i dZ_i) - \Delta_i M(\mu_M dt + \sigma_M dZ_M) \\
&= \mu_i S_i dt + \sigma_i S_i(\rho_{iM} dZ_M + \sqrt{1 - \rho_{iM}^2}\, \varepsilon_i) - \Delta_i M(\mu_M dt + \sigma_M dZ_M) \\
&= (\mu_i S_i - \Delta_i \mu_M M)dt + (\rho_{iM}\sigma_i S_i - \Delta_i \sigma_M M)dZ_M + \sigma_i S_i \sqrt{1 - \rho_{iM}^2}\, \varepsilon_i
\end{aligned}
\qquad [2.28]
$$

We can eliminate all of the risk of $\tilde{S}_i$ with respect to market moves $dZ_m$ by choosing $\rho_{iM}\sigma_i S_i - \Delta_i \sigma_M M = 0$, so that the short position in M at any instant is given by

$$\Delta_i = \frac{\rho_{iM}\sigma_i S_i}{\sigma_M M} = \frac{\rho_{iM}\sigma_i \sigma_M S_i}{\sigma_M^2 M} = \beta_{im}\frac{S_i}{M} \qquad [2.29]$$

where

$$\beta_{im} = \frac{\rho_{iM}\sigma_i \sigma_M}{\sigma_M^2} = \frac{\sigma_{iM}}{\sigma_M^2} \qquad [2.30]$$

is the traditional beta, the ratio of the covariance $\sigma_{iM}$ of stock $i$ with the market to the variance of the market $\sigma_M^2$.

By substituting the value of $\Delta_i$ in Equation 2.29 into Equation 2.27 one finds that the value of the market-neutral version of $S_i$ is

$$\tilde{S}_i = (1 - \beta_{iM})S_i \qquad [2.31]$$

By using the same value of $\Delta_i$ in the last line of Equation 2.28 one can write the evolution of $\tilde{S}_i$ as

$$\frac{d\tilde{S}_i}{\tilde{S}_i} = \tilde{\mu}_i dt + \tilde{\sigma}_i \varepsilon_i \qquad [2.32]$$

where

$$\tilde{\mu}_i = \frac{\mu_i - \beta_{iM}\mu_M}{1 - \beta_{iM}}$$

$$\tilde{\sigma}_i = \frac{\sigma_i\sqrt{1 - \rho_{iM}^2}}{1 - \beta_{iM}} \qquad [2.33]$$

These equations describe the stochastic evolution of the market-hedged component of stock *i*, its expected return and volatility modified by the hedging of market-correlated movements.

The evolution of the hedged components of two different stocks S and P is described by

$$\frac{d\tilde{S}}{\tilde{S}} = \tilde{\mu}_S \, dt + \tilde{\sigma}_S \varepsilon_S$$

$$\frac{d\tilde{P}}{\tilde{P}} = \tilde{\mu}_P dt + \tilde{\sigma}_P \varepsilon_P \qquad [2.34]$$

What is the relation between the expected returns of these two hedged portfolios?

Again, assuming $\tilde{\sigma}_P > \tilde{\sigma}_S$, we can at any instant create a portfolio V consisting of *w* shares of $\tilde{P}$ and $1 - w$ shares of the riskless bond B, with *w* chosen so that the volatility of V is instantaneously the same as that of $\tilde{S}$. Then, according to our invariance principal, V and $\tilde{S}$ must have the same expected return. More succinctly, if $\sigma_V = \tilde{\sigma}_S$, then $\mu_V = \tilde{\mu}_S$.

Repeating the algebraic arguments that led to Equation 2.12, we obtain the constraint

$$\frac{\tilde{\mu}_S - r}{\tilde{\sigma}_S} = \frac{\tilde{\mu}_P - r}{\tilde{\sigma}_P} = \lambda$$

Substitution of Equation 2.33 for $\tilde{\mu}_i$ and $\tilde{\sigma}_i$ leads to the result

$$(\mu_S - r) - \beta_{SM}(\mu_M - r) = \lambda\sigma_S\sqrt{1 - \rho_{SM}^2} \qquad [2.35]$$

Equation 2.35 shows that if you can hedge away the market component of any stock S, its excess return less $\beta_{SM}$ times the excess return of the market is proportional to the component of the volatility of the stock orthogonal to the market.

## 2.6 Diversifiable Stocks Correlated with One Market Factor

We now repeat the arguments of Section 2.4 in the case where one can diversify the non-market risk over a portfolio V consisting of L stocks whose residual movements are on average uncorrelated and whose variance $\sigma_V$ is therefore $O(1/L)$ as $L \to \infty$.

If our invariance principle is to apply to portfolios of stocks, then Equation 2.35 must hold for V, so that

$$(\mu_V - r) - \beta_{VM}(\mu_M - r) \sim \lambda \sigma_V \sqrt{1 - \rho_{VM}^2} \sim 0 \,.$$

where the right hand side of the above relation is asymptotically zero because $\sigma_V \to 0$.

By decomposing the zero-variance portfolio V into its constituents, we can analogously repeat the argument that led from Equation 2.21 to Equation 2.22 to show that $\lambda \sim 0$. Therefore, Equation 2.35 reduces to

$$(\mu_S - r) = \beta_{SM}(\mu_M - r) \qquad\qquad [2.36]$$

This is the well-known result of the Capital Asset Pricing Model, which states that the excess expected return of a stock is related beta times the excess return of the market.

# 3 The Invariance Principle in Intrinsic Time

## 3.1 Trading Frequency, Speculation and Intrinsic Time

Investors are generally accustomed to evaluating the returns they can earn and the volatilities they will experience with respect to some interval of calendar time, the time continuously measured by a standard clock, common to all investors and markets. The passage of calendar time is unaffected by and unrelated to the vagaries of trading in a particular stock.

However, stocks do not trade continuously; each stock has its own trading patterns. Stocks trade at discrete times, in finite amounts, in quantities constrained by supply and demand. The number of trades and the number of shares traded per unit of time both change from minute to minute, from day to day and from year to year. Opportunities to profit from trading depend on the amount of stock available and the trading frequency.

Over the long run, over years or months or perhaps even weeks, opportunities average out. In the end, people live their lives and work at their jobs and build their companies in calendar time. Therefore, for most stocks and markets, for most of the time, there is little relationship between the frequency of trading opportunities and expected risk and return. The bond market's expected returns are particularly likely to be insensitive to trading frequency, since, unlike stocks, a bond's coupons and yields are contractually specified in terms of calendar time.

Nevertheless, in highly speculative and rapidly developing market sectors where relevant news arrives frequently, expectations can suddenly soar and investors may have very short-term horizons. The internet sector, communications and biotechnology are recent examples. In markets such as these there may be a psychological interplay between high trading frequency and expected return. This sort of inter-relation could take several forms.

On the simplest and most emotional level, speculative excitement coupled to the expectation of outsize returns can lead to a higher frequency of trading. But, more subtly, investors or speculators with very short-term horizons may apprehend risk and return differently. Day-traders may instinctively prefer to think of a security's risk and return as being characterized by the time intervals between the passage of trading opportunities.

Each stock has its own intrinsic rate for the arrival of trading opportunities. There is a characteristic minimum time for which a trade must be held, a minimum time before it can be unwound. Short-term speculators may rationally choose to evaluate the relative merits of competing investments in terms of the risk and return they promise over one trading interval.

We refer, somewhat loosely for now, to the frequency of trading opportunities in calendar time as the stock's *trading frequency*. One way of thinking about it is as the number of trades occurring per day. The trading frequency $v_i$ of a stock has the dimension $[\text{time}]^{-1}$, and therefore implicitly determines an *intrinsic-time*[11] scale $\tau_i$ for that stock, a time ticked off by an imaginary clock that measures the passing of opportunities for trading that stock. This trading frequency determines a linear mapping between the stock's intrinsic time $\tau_i$ and standard calendar time $t$. We will define and discuss these relationships more carefully in the following sections. Trading frequencies vary with time in both systematic and random ways, but, in the models of this paper, we will focus only the *average* trading frequency of the stock, and ignore the effects of its fluctuations.

---

11. See [Müller 1995].

When one compares speculative short-term investments in several securities, one must be aware that the minimum calendar-time interval between definable trading opportunities differs from security to security. For example, riskless interest-rate investments typically occur overnight, and therefore have a minimum scale of about one day. In contrast, S&P 500 futures trades may have a time scale of minutes or hours. These intervals between effective opportunities, vastly different in calendar time, represent the same amount of a more general trading-opportunity or intrinsic time.

We have been purposefully vague in specifying exactly what is meant by a trading opportunity. In a model of market microstructure it would be determined by the way agents behave and respond. In this paper it is closer to an effective variable that represents the speed or liquidity of a market, There are several possible ways, listed below, in which trading opportunities and the time interval between them can be quantified, each with different economic meanings:

1. The simplest possibility is to imagine a trading opportunity as the chance to perform a trade, independent of size. The reciprocal of the time interval between trades is the least complex notion of trading frequency. In this view, a high trading frequency corresponds simply to rapid trading.

2. A second possibility is to interpret a trading opportunity as the chance to trade *a fixed number* of shares. The time interval between trades is then a measure of the average time elapsed per share traded. Here a high trading frequency corresponds to high liquidity.

3. A third alternative is to regard a trading opportunity as the chance to trade *a fixed percentage* of the float for that stock. The reciprocal of the stock's trading frequency then measures the average time elapsed per some percentage of the float traded. In this view a high trading frequency means that large fractions of the available float trade in a short amount of (calendar) time. This means that not much excess stock is available, making the stock relatively illiquid.

4. Another possibility is to think of the time interval between trading opportunities as the average time between the arrival of bits of company-specific information.

It is not obvious which of these alternatives is to be preferred. It is likely that different markets may see significance in different definitions of trading opportunity. In the end, the trading frequency for a specific stock may best be regarded as an implied variable, its value to be inferred from market features that depend on it.

We now proceed to investigate the consequences of the hypotheses that (1) each security has its own intrinsic time scale, and (2) that some investors, especially short-term speculators, care about the relative risk and return of securities as perceived and measured in this intrinsic time.

## 3.2 The Definition of Intrinsic Time

We begin by assuming that short-term investors perceive a stock's price to evolve as a function of the time interval between trading opportunities. We therefore replace Equation 2.1 by

$$\frac{dS_i}{S_i} = M_i d\tau_i + \Sigma_i dW_i \qquad [3.1]$$

Here, $d\tau_i$ represents an infinitesimal increment in the *intrinsic time* $\tau_i$ that measures the rate at which trading opportunities for stock $i$ pass. The symbol $M_i$ represents the expected return of stock *i per unit of its intrinsic time* and $\Sigma_i$ denotes the stock's volatility *measured in intrinsic time*, as given by the square root of the variance of the stock's returns per unit of intrinsic time.

Analogous to Equation 2.2, we write

$$dW_i^2 = d\tau_i$$
$$dW_i dW_j = \pi_{ij}\sqrt{d\tau_i}\sqrt{d\tau_j} \qquad [3.2]$$

where $\pi_{i,j}$ is the correlation between the intrinsic-time returns of stock $i$ and stock $j$.

By the *intrinsic time* of a stock we mean the time measured by a single, universal conceptual clock which ticks off *one unit (a tick, say) of intrinsic time* with the passage of each successive "trading opportunity" for that stock. Intrinsic time is *dimensionless*; it simply counts the passage of trading opportunities.

The ratio between a tick of intrinsic time and a second of calendar time varies from stock to stock, depending on the rate at which each stock's trading opportunities occur. Even for a single stock, the ratio between a tick and a second changes from moment to moment. In the models developed in this paper, for each stock we will focus only on the average ratio of a tick to a second, and use that average ratio to define the trading frequency for the stock. We will ignore the effects of fluctuations in the ratio.

The notion that, during certain periods, a stock's price evolves at a pace determined by its own intrinsic clock is not necessarily that strange. Stock price changes are triggered by news or noise, both of whose rates of arrival differ from industry to industry. For new industries still in the process of being evaluated and re-evaluated as product development, consumer acceptance and competitor response play leapfrog with each other, the intrinsic time of stock evolution may pass more rapidly than it does for mature industries. Newly developing markets can burn brightly, passing from birth to death in one day. Dull and routine industries can slumber fitfully for long periods.

If investors' measures of risk and return are intuitively formulated in intrinsic time, we must relate their description to the market's commonly quoted measures of risk and return in calendar time.

## 3.3 Converting from Intrinsic to Calendar Time

Investors commonly speak about risk, correlation and return as measured in calendar time. In Equations 2.1, and 2.2, $\mu_i$ denotes the expected return (per calendar day, for example), $\sigma_i$ denotes the volatility of these calendar-time returns, and $\rho_{ij}$ is their correlation.

Suppose that, intuitively, investors "think" about a stock's future evolution in intrinsic time, as described by Equation 3.1, where $M_i$ denotes the expected return of stock *i per tick of intrinsic time,* $\Sigma_i$ denotes the volatility of these intrinsic-time returns, and $\pi_{ij}$ is their correlation. What is the relationship between the intrinsic-time and calendar-time measures?

We define a stock $i$'s *trading frequency* $\nu_i$ to be the number of intrinsic-time ticks that occur for the stock $i$ in one calendar second. The higher the trading frequency $\nu_i$ for a stock $i$, the more trading opportunities pass by per calendar second. The relationship between the flow of calendar time $t$ and the flow of intrinsic time $\tau_i$ is given by

$$d\tau_i = \nu_i dt \qquad\qquad [3.3]$$

This relationship differs from stock to stock, varying with each stock's trading frequency. Although actual trading frequencies vary from second to second, we stress again that, in this paper, we make the approximation that $\nu_i$ for each stock is constant through time.

It is customary to think of calendar time $t$ as a universal, stock-independent measure; nevertheless, for the remainder of this paper, it will be convenient to think of intrinsic time $\tau$ as the universal quantity, the measure which counts the interval between any two successive ticks of any stock as counted as one universal unit. Since intrinsic time is dimensionless and merely counts the evolution of trading opportunities, the dimensionality of $\nu_i$ is [time]$^{-1}$.

$M_i$ in Equation 3.1 is the stock's return per tick. Therefore, the stock's return in one calendar second consisting of $\nu_i$ ticks is given by

$$\mu_i = \nu_i M_i \qquad\qquad [3.4]$$

$\Sigma_i$ in Equations 3.1 is the volatility of the intrinsic-time returns. The volatility in calendar time is given by

$$\sigma_i = \sqrt{\nu_i}\Sigma_i \qquad\qquad [3.5]$$

where the square root is the familiar consequence of the additivity of variance for independent random variables.

The relationship between the intrinsic-time correlation $\pi_{ij}$ and calendar-time correlation $\rho_{ij}$ is simpler: since they are both dimensionless, they are identical. One can show this by using Equation 3.1 to write

$$\frac{dS_i}{S_i}\frac{dS_j}{S_j} = \Sigma_i\Sigma_j dW_i dW_j = \Sigma_i\Sigma_j\pi_{ij}\sqrt{d\tau_i}\sqrt{d\tau_j} = \Sigma_i\Sigma_j\pi_{ij}dt\sqrt{v_i}\sqrt{v_j}$$

where the last equality follows from Equation 3.3. However, from Equations 2.1 and 2.2, one can also write

$$\frac{dS_i}{S_i}\frac{dS_j}{S_j} = \rho_{ij}\sigma_i\sigma_j dt = \rho_{ij}\sqrt{v_i}\Sigma_i\sqrt{v_i}\Sigma_j dt$$

where the last equality follows from Equation 3.5. Comparing the above two expressions, we see that

$$\pi_{ij} = \rho_{ij} \qquad [3.6]$$

In deriving this result we have again assumed that the trading frequencies are not stochastic.

In terms of the familiarly quoted calendar-time risk variables, Equation 3.1 can be re-expressed as the intrinsic-time Wiener process

$$\frac{dS_i}{S_i} = \frac{\mu_i}{v_i}d\tau_i + \frac{\sigma_i}{\sqrt{v_i}}dW_i \qquad [3.7]$$

Note that when compared with the calendar-time evolution of Equation 2.1, the expected returns $\mu_i$ are scaled by $v_i$ and the volatilities $\sigma_i$ are scaled by $\sqrt{v_i}$, as must be the case on dimensional grounds, since $\tau_i$ and $W_i$ are dimensionless.

## 3.4 The Invariance Principle in Intrinsic Time

We now begin to explore the consequence of the simple invariance principle of Section 2.2, modifying it so that the risk and return it refers to are measured in intrinsic time. In this form, the principle states that

> *Two portfolios with the same perceived intrinsic-time risk should have the same expected intrinsic-time return.*

Of course, the respective calendar-time intervals over which these two identical returns are expected to be realized are not equal to each other, but are related through the ratio of their trading frequencies.

## 3.5 Living in Intrinsic Time

Henceforth, we want to take the view of someone who wears an intrinsic-time wristwatch and cares only about the number of ticks that pass. For him or her, the amount of calendar time between ticks is irrelevant. What matters is the risk and return per tick, and all ticks, no matter how long the interval between them in calendar time, are equivalent. From now on, we assume that intrinsic time, rather than calendar time, is the universal measure.

We can then replace all security-specific intrinsic time scales $\tau_i$ by a single $\tau$ scale that simply counts ticks. Equation 3.8 for the perceived evolution of any stock $i$ can be rewritten as the Wiener process

$$\frac{dS_i}{S_i} = \frac{\mu_i}{\nu_i}d\tau + \frac{\sigma_i}{\sqrt{\nu_i}}dW_i \qquad [3.8]$$

where

$$dW_i^2 = d\tau$$
$$dW_i dW_j = \rho_{ij}d\tau \qquad [3.9]$$

The calendar-time stock evolution of Equation 2.1 is related to the intrinsic-time evolution of Equation 3.8 by following simple transformation:

$$t \to \tau$$
$$\mu_i \to \frac{\mu_i}{\nu_i} \qquad [3.10]$$
$$\sigma_i \to \frac{\sigma_i}{\sqrt{\nu_i}}$$

These $\nu_i$-dependent scale factors provide the only dimensionally consistent conversion from $t$- to $\tau$-evolution, since $\tau_i$ and $W_i$ in Equation 3.8 are dimensionless.

## 3.6 A Digression on the Comparison of One-Tick Investments

As long as one uses the $\tau$ scale to think in intrinsic time, all our previous invariance arguments for portfolios will be easy to duplicate. This is the path we will take, beginning in Section 3.7. But, if every security marches to the beat of its own drum, what investment scenario in calendar time is one actually contemplating when one thinks about the risk and return of a multi-asset portfolio on the $\tau$ scale? Here we provide a brief account of what it means to compare the results of one-tick-long investments.

A tick, the reciprocal of the trading frequency $\nu_i$, is the shortest possible holding time for an investment in a security $i$. The intrinsic-time viewpoint regards each security as being held for just one finite-length tick, even though each security's tick length differs from another's when expressed in calendar time. However, the profit or loss from a one-tick-long investment cannot be realized immediately. The current conventions of trade settlement require waiting at least one full day to realize the proceeds of an intra-day trade.

Consider a riskless bond. As pointed out earlier, the guaranteed returns on bonds are inextricably bound to calendar time; bonds pay interest and principal on definite calendar dates. In fixed-income markets, the shortest period over which one can earn guaranteed and riskless interest is overnight. The trading frequency $v_B$ of a riskless bond B is therefore about once per day, much longer than the typical stock tick length. Although one can formally write the continuous differential equation for the price of a riskless bond as

$$\frac{dB}{B} = r\,dt \qquad\qquad [3.11]$$

$dt$ is not strictly an infinitesimal. The bond's evolution in intrinsic time is found by combining Equation 3.3 with Equation 3.11 to obtain

$$\frac{dB}{B} = \frac{r}{v_B}\,d\tau. \qquad\qquad [3.12]$$

Equation 3.12 should not be interpreted to mean that a riskless bond can earn a fraction $(r/v_B)$ of its daily interest $r$ during an infinitesimal time $d\tau$. Instead, it means that if you hold the stock for the minimum time of one tick, about a day long, you will earn interest $r$. There is no shorter investment period than $(1/v_B)$.

Now consider a one-tick-long investment in a portfolio containing stocks $S_i$ with corresponding trading frequency $v_i$. Stocks require at least one day to settle. A speculator who buys a stock and then quickly sells it a tick or two later does not receive the proceeds, or begin to earn any interest from their riskless reinvestment, until at least the start of the next day, No matter how long each stock's tick, the resultant profit or loss on all the stocks in the portfolio, each held for one tick, can only be realized a day later, when all the trades have settled.

Equation 3.8 describes the perceived evolution of stocks in a portfolio, each of which is held for one intrinsic tick and then unwound, with the return being evaluated a day later, where one day is the tick length of the riskless bond investment which provides the benchmark return. More generally, the portfolio member with the lowest trading frequency determines the shortest holding time after which all results can be evaluated.

One last point: the daily volatility of a position in speculative stocks is commonly large enough to cause price moves much greater than the amount of interest to be earned from a corresponding position in riskless bonds. Therefore, it will often not be a bad approximation so simply set $r$ equal to zero in order to derive simple approximate risk-return relations for short-term speculative trades.

### 3.7 Uncorrelated Stocks in an Undiversifiable Market: The Intrinsic-Time Sharpe Ratio

We now begin to parallel the arguments of Section 2.3, modifying them to take account of risk and return as perceived from an intrinsic-time point of view.

Consider two stocks $S$ and $P$ whose perceived short-term evolution is described by the intrinsic-time Wiener process of Equation 3.8:

$$\frac{dS}{S} = \frac{\mu_S}{v_S}d\tau + \frac{\sigma_S}{\sqrt{v_S}}dW_S$$

$$\frac{dP}{P} = \frac{\mu_P}{v_P}d\tau + \frac{\sigma_P}{\sqrt{v_P}}dW_P$$

[3.13]

For each stock $i$, $v_i$ is its trading frequency, $\mu_i$ its expected return *in calendar time*, and $\sigma_i$ the volatility of its calendar-time returns We assume that $\sigma_P/(\sqrt{v_P})$ is greater than $\sigma_S/(\sqrt{v_S})$.

Given Equations 3.12 and 3.13, what is the appropriate relationship between the expected returns $\mu_S$ and $\mu_P$? We can repeat the arguments of Section 2.3, now using the invariance principle as interpreted in intrinsic time to derive parallel formulas by respectively replacing $\mu_i$ by $\mu_i/v_i$ and $\sigma_i$ by $\sigma_i/(\sqrt{v_i})$, as indicated by the transformation of Equation 3.10.

As before, we construct a portfolio V that is less risky than P by adding to it some amount of the riskless bond B, so that

$$V = wP + (1-w)B$$

[3.14]

From Equations 3.12 and 3.13, the evolution of V during time $d\tau$ is described by

$$\frac{dV}{V} = M_V d\tau + \Sigma_V dW_P$$

and

$$M_V = \frac{w(\mu_P/v_P)P + (1-w)(r/v_B)B}{wP + (1-w)B}$$

$$\Sigma_V = \frac{wP(\sigma_P/\sqrt{v_P})}{wP + (1-w)B}$$

[3.15]

The intrinsic-time invariance principle demands that equal risk produce equal expected return. As before, we require that when $w$ is chosen to give V and S the same volatility in intrinsic time, then V and S must also have the same expected return per unit of intrinsic time.

The value of $w$ that guarantees that $\Sigma_V = \Sigma_S \equiv \dfrac{\sigma_S}{\sqrt{v_S}}$ is given by

$$\frac{1}{w} = 1 + \frac{P}{B}\left(\frac{\sigma_P}{\sigma_S}\sqrt{\frac{v_S}{v_P}} - 1\right) \qquad [3.16]$$

The value of $w$ that guarantees that $M_V = M_S \equiv \dfrac{\mu_S}{v_s}$ is given by

$$\frac{1}{w} = 1 + \frac{P}{B}\frac{\left(\dfrac{\mu_P}{v_P} - \dfrac{\mu_S}{v_s}\right)}{\left(\dfrac{\mu_S}{v_s} - \dfrac{r}{v_B}\right)} \qquad [3.17]$$

Equations 3.16 and 3.17 are consistent only if

$$\frac{(\mu_S / v_S) - (r / v_B)}{\sigma_S / (\sqrt{v_S})} = \frac{(\mu_P / v_P) - (r / v_B)}{\sigma_P / (\sqrt{v_P})} \qquad [3.18]$$

Therefore, analogously to the argument leading to Equation 2.13, we conclude that for any stock $i$

$$\frac{(\mu_i / v_i) - (r / v_B)}{\sigma_i / (\sqrt{v_i})} = \Lambda \qquad [3.19]$$

where $\Lambda$, the intrinsic-time Sharpe ratio, is the analog of the standard Sharpe ratio, and is dimensionless.

Equation 3.19 is a short-term, trading-frequency-sensitive version of the risk-return relation of Equation 2.13 that can only hold over relatively brief time periods, since, in the long run, the ultimate performance of a company cannot depend on the frequency at which its stock is traded.

## 3.8 A Stock's Temperature

We can rewrite Equation 3.19 in the form

$$\mu_i - r(v_i / v_B) = \Lambda \sigma_i \sqrt{v_i} \qquad [3.20]$$

First, imagine that the riskless rate $r$ is zero. Then, Equation 3.20 reduces to $\mu_i = \Lambda \sigma_i \sqrt{v_i}$, which states that the expected return on any stock is proportional to the product of its (calendar-time) volatility and the square root of its trading frequency.

For brevity, we will refer to the quantity

$$\chi_i = \sigma_i \sqrt{v_i} \qquad [3.21]$$

as the *temperature* of the stock. It provides a measure of the perceived speculative riskiness of the stock in terms of how it influences expected return. Since both $\sigma_i$ and $\sqrt{v_i}$ have dimension [seconds]$^{-1/2}$, temperature has the dimension [seconds]$^{-1}$, and, as stressed before, $\Lambda$ is dimensionless. For a market of undiversifiable stocks, Equation 3.20 states that expected return is proportional to temperature. In terms of intrinsic-time volatility $\Sigma_i$, the temperature can also be written as

$$\chi_i = \Sigma_i v_i \qquad [3.22]$$

thereby demonstrating the role that both volatility and frequency play in determining perceived risk.

Let us define the *frequency-adjusted riskless rate $R_i$* to be

$$R_i = r \frac{v_i}{v_B} \qquad [3.23]$$

$R_i$ is the riskless rate scaled by the ratio of the trading frequency of the stock to that of the riskless bond. It represents the rate at which interest would be earned if the riskless bond's trading frequency were the same as that of the stock.

In terms of these variables, Equation 3.20 can be rewritten as

$$\frac{\mu_i - R_i}{\chi_i} = \Lambda \qquad [3.24]$$

It states that for each stock, the expected return in excess of the frequency-adjusted riskless rate per unit of temperature is the same for all stocks[12]. Note that both the frequency-adjusted riskless rate (relative to which excess return is measured) and the temperature (which determines the risk responsible for the excess return) increase monotonically with trading frequency $v_i$.

Nothing yet tells us the value of the intrinsic-time Sharpe ratio $\Lambda$.

---

12. This equation for $\Lambda$ resembles the definition of entropy in thermodynamics. To the extent that one can identify excess return with the rate of heat flow from a hot source and $\chi_i$ as the temperature at which the flow takes place, $\Lambda$ then corresponds to the rate of change of entropy as stock prices grow.

## 3.9 Uncorrelated Stocks in a Diversifiable Market

We again duplicate the arguments of Section 2.4, modifying them for short-term trades whose risk is perceived from an intrinsic-time point of view

As before, consider a portfolio V consisting of $l_i$ shares of each of $L$ different stocks, whose value is given by

$$V = \sum_{i=1}^{L} l_i S_i \qquad [3.25]$$

Then the change in value over one infinitesimal increment of intrinsic time $d\tau$ is given by

$$dV = \sum_{i=1}^{L} l_i dS_i = \sum_{i=1}^{L} l_i S_i \left( \frac{\mu_i}{\nu_i} d\tau + \frac{\sigma_i}{\sqrt{\nu_i}} dW_i \right)$$

$$= \left( \sum_{i=1}^{L} l_i S_i \frac{\mu_i}{\nu_i} \right) d\tau + \sum_{i=1}^{L} l_i S_i \frac{\sigma_i}{\sqrt{\nu_i}} dW_i$$

The instantaneous return on this portfolio is

$$\frac{dV}{V} = \left( \sum_{i=1}^{L} w_i \frac{\mu_i}{\nu_i} \right) d\tau + \sum_{i=1}^{L} w_i \frac{\sigma_i}{\sqrt{\nu_i}} dW_i \qquad [3.26]$$

where the fixed weights $w_i$ are given, as before, by Equation 2.17.

From Equation 3.26, the expected return of V per unit of intrinsic time is given by

$$M_V = \sum_{i=1}^{L} w_i \frac{\mu_i}{\nu_i} \qquad [3.27]$$

The variance of these returns is

$$\Sigma_V^2 = \sum_{i,j=1}^{L} w_i w_j \rho_{ij} \frac{\sigma_i \sigma_j}{\sqrt{\nu_i \nu_j}}$$

$$= \sum_{i=1}^{L} w_i^2 \frac{\sigma_i^2}{\nu_i} + \sum_{i \neq j} w_i w_j \rho_{ij} \frac{\sigma_i \sigma_j}{\sqrt{\nu_i \nu_j}} \qquad [3.28]$$

As in Section 2.4, if all stocks are approximately equally weighted so that $w_i \sim O(1/L)$, and if, on average, their returns are uncorrelated with each other, so that $\rho_{ij} < 0(1/L)$, then

$$\Sigma_V^2 \sim O(1/L) \to 0 \quad \text{as} \quad L \to \infty \qquad [3.29]$$

Asymptotically, as $L \to \infty$, the variance of the intrinsic-time returns of the portfolio approaches zero and V becomes riskless.

We now apply Equation 3.24 to the entire portfolio V. Since V is riskless and $\Sigma_V$ is zero, it follows from Equation 3.22 that

$$\mu_V - R_V = \Lambda \chi_V = \Lambda \nu_V \Sigma_v = 0$$

and so

$$\mu_V = R_V. \qquad [3.30]$$

But, since $\mu_V = \nu_V M_V$ and $R_V = r \dfrac{\nu_V}{\nu_B}$, Equation 3.30 implies that

$$M_V = \frac{r}{\nu_B} \qquad [3.31]$$

Therefore, the intrinsic-time expected return of portfolio V is just the intrinsic time expected return of the riskless bond.

We can now also substitute from Equation 3.27 and Equation 3.24 to write

$$M_V = \sum_{i=1}^{L} w_i \frac{\mu_i}{\nu_i} = \sum_{i=1}^{L} w_i \frac{R_i + \Lambda \chi_i}{\nu_i} = \sum_{i=1}^{L} w_i \frac{r(\nu_i / \nu_B) + \Lambda \chi_i}{\nu_i} .$$
$$= \frac{r}{\nu_B} + \Lambda \sum_{i=1}^{L} \frac{w_i \chi_i}{\nu_i}$$

Comparing this last equation with Equation 3.31 we see that $\Lambda = 0$.

We conclude that, in order that the excess return per degree of temperature be the same for any stock as well as for a diversified portfolio, the intrinsic-time Sharpe ratio must be zero. Therefore, from Equation 3.24, the expected return in calendar time for stocks in a diversifiable universe satisfies

$$\mu_i = r(\nu_i / \nu_B),$$

and is equal to the frequency-adjusted riskless rate.

## 3.10 Undiversifiable Stocks Correlated with One Market Factor

We now turn to the risk-return relationship for a market in which all stocks are correlated with one market factor and in which perceptions are based on intrinsic time. We parallel the derivation of Section 2.5, assuming that the market consists of a riskless bond B and N companies, each company $i$ having issued $n_i$ stocks of current market value $S_i$. We also assume the existence of a traded index M that represents the entire market.

The evolution of these securities satisfies

$$\frac{dM}{M} = \frac{\mu_M}{v_M}d\tau + \frac{\sigma_M}{\sqrt{v_M}}dW_M$$

$$\frac{dS_i}{S_i} = \frac{\mu_i}{v_i}d\tau + \frac{\sigma_i}{\sqrt{v_i}}dW_i \qquad [3.32]$$

$$\frac{dB}{B} = \frac{r}{v_B}d\tau$$

The correlation of each stock with the market factor M is given by

$$dW_i = \rho_{iM}dW_M + \sqrt{1 - \rho_{iM}^2}\,\varepsilon_i \qquad [3.33]$$

Here $\varepsilon_i$ is a random normal variable that represents the idiosyncratic risk of stick $i$, and is uncorrelated with $dW_M$. Perceiving risk in intrinsic time, we also assume that

$$\varepsilon_i^2 = dW_M^2 = dW_i^2 = d\tau$$

and

$$dW_i dW_M = \rho_{iM}d\tau.$$

As before, let

$$\tilde{S}_i = S_i - \Delta_i M \qquad [3.34]$$

denote the value of the market-neutral portfolio corresponding to the stock $S_i$. We can now duplicate the arguments of Section 2.5, thereby deriving similar formulas to those that appear there by respectively replacing $\mu_i$ by $\mu_i/v_i$ and $\sigma_i$ by $\sigma_i/(\sqrt{v_i})$.

The evolution of $\tilde{S}_i$ in intrinsic time is given by

$$\begin{aligned}
d\tilde{S}_i &= dS_i - \Delta_i dM \\
&= S_i\left(\frac{\mu_i}{v_i}d\tau + \frac{\sigma_i}{\sqrt{v_i}}dW_i\right) - \Delta_i M\left(\frac{\mu_M}{v_M}d\tau + \frac{\sigma_M}{\sqrt{v_M}}dW_M\right) \\
&= \frac{\mu_i}{v_i}S_i d\tau + \frac{\sigma_i}{\sqrt{v_i}}S_i(\rho_{iM}dW_M + \sqrt{1-\rho_{iM}^2}\,\varepsilon_i) - \Delta_i M\left(\frac{\mu_M}{v_M}dt + \frac{\sigma_M}{\sqrt{v_M}}dW_M\right) \\
&= \left(\frac{\mu_i}{v_i}S_i - \Delta_i\frac{\mu_M}{v_M}M\right)dt + \left(\rho_{iM}\frac{\sigma_i}{\sqrt{v_i}}S_i - \Delta_i\frac{\sigma_M}{\sqrt{v_M}}M\right)dW_M + \frac{\sigma_i}{\sqrt{v_i}}S_i\sqrt{1-\rho_{iM}^2}\,\varepsilon_i
\end{aligned} \qquad [3.35]$$

We can eliminate all risk with respect to market-index moves $dW_M$ by choosing $\rho_{iM}\dfrac{\sigma_i}{\sqrt{\nu_i}}S_i - \Delta_i\dfrac{\sigma_M}{\sqrt{\nu_M}}M = 0$, so that, solving for $\Delta_i$, we obtain

$$\Delta_i = \frac{\rho_{iM}\sigma_i S_i}{\sigma_M M}\sqrt{\frac{\nu_M}{\nu_i}} = \frac{\rho_{iM}\sigma_i\sigma_M S_i}{\sigma_M^2 M}\sqrt{\frac{\nu_M}{\nu_i}} = \beta_{iM}\sqrt{\frac{\nu_M}{\nu_i}}\frac{S_i}{M} \qquad [3.36]$$

where

$$\beta_{im} = \frac{\rho_{iM}\sigma_i\sigma_M}{\sigma_M^2} \qquad [3.37]$$

is the familiar beta of the Capital Asset Pricing Model.

By substituting this value of $\Delta_i$ into Equation 3.34 one finds that

$$\tilde{S}_i = \left(1 - \beta_{iM}\sqrt{\frac{\nu_M}{\nu_i}}\right)S_i \qquad [3.38]$$

By using the same value of $\Delta_i$ in the last line of Equation 3.35 one can write the evolution of $\tilde{S}_i$ as

$$\frac{d\tilde{S}_i}{\tilde{S}_i} = \frac{\tilde{\mu}_i}{\nu_i}d\tau + \frac{\tilde{\sigma}_i}{\sqrt{\nu_i}}\varepsilon_i \qquad [3.39]$$

where the expected return and the volatility are given by

$$\tilde{\mu}_i = \frac{\mu_i - \beta_{iM}\mu_M\sqrt{\dfrac{\nu_i}{\nu_M}}}{\left(1 - \beta_{iM}\sqrt{\dfrac{\nu_M}{\nu_i}}\right)}$$

$$\qquad [3.40]$$

$$\tilde{\sigma}_i = \frac{\sigma_i\sqrt{1 - \rho_{iM}^2}}{\left(1 - \beta_{iM}\sqrt{\dfrac{\nu_M}{\nu_i}}\right)}$$

These equations describe the stochastic intrinsic-time evolution of the market-hedged component of stock $i$.

Now consider two market-hedged stocks S and P that evolve according to

$$\frac{d\tilde{S}}{\tilde{S}} = \frac{\tilde{\mu}_S}{v_S}d\tau + \frac{\tilde{\sigma}_S}{\sqrt{v_S}}\varepsilon_S$$

$$\frac{d\tilde{P}}{\tilde{P}} = \frac{\tilde{\mu}_P}{v_P}d\tau + \frac{\tilde{\sigma}_P}{\sqrt{v_P}}\varepsilon_P$$

[3.41]

Assuming $\frac{\tilde{\sigma}_P}{\sqrt{v_P}} > \frac{\tilde{\sigma}_S}{\sqrt{v_S}}$, we can again create a portfolio V consisting of $w$ shares of $\tilde{P}$ and $1 - w$ shares of the riskless bond B, choosing $w$ so that V and S have the same intrinsic-time risk, and, therefore, the same intrinsic-time expected return. Repeating the argument that led to Equation 3.19, we obtain the risk-reward relation

$$\frac{\frac{\tilde{\mu}_S}{v_S} - \frac{r}{v_B}}{\frac{\tilde{\sigma}_S}{\sqrt{v_S}}} = \Lambda$$

Using Equation 3.40 to expand $\tilde{\mu}_S$ and $\tilde{\sigma}_S$ leads to the result

$$\left(\frac{\mu_S}{v_S} - \frac{r}{v_B}\right) - \beta_{SM}\sqrt{\frac{v_M}{v_S}}\left(\frac{\mu_M}{v_M} - \frac{r}{v_B}\right) = \Lambda\frac{\sigma_S}{\sqrt{v_S}}\sqrt{1 - \rho_{SM}^2}$$

[3.42]

This equation relates the excess return of a non-diversifiable stock to the volatility and correlation of the stock itself. $\Lambda$ is the intrinsic-time Sharpe ratio, dimensionless and of unknown value.

In order to obtain a little more intuition about this equation, we examine it for very short-term trades during which negligible interest is earned. Setting the interest rate $r = 0$, we obtain the approximate formula

$$\mu_S - \beta_{SM}\sqrt{\frac{v_S}{v_M}}\mu_M = \Lambda\chi_S\sqrt{1 - \rho_{SM}^2}$$

[3.43]

where $\chi_S$ is the trading temperature of stock S as defined in Equation 3.21. The left-hand-side of this formula suggests that, for short-term speculators who think about stocks from an intrinsic-time point of view, the benchmark return is beta times the market return, *enhanced by a factor equal to the square root of the ratio of the trading frequency of the stock to that of the market.* Furthermore, the residual return above this benchmark is proportional to the stock's trading temperature rather than simply its volatility. These frequency-dependent factors can be appreciable for stocks whose trading frequency is high relative to that of the market.

### 3.11 Diversifiable Stocks Correlated with One Market Factor: The Intrinsic-Time Capital Asset Pricing Model

We now parallel the arguments of Section 2.6 in the case where one can diversify into a market-hedged multi-stock portfolio V whose intrinsic-time residual volatility asymptotically becomes zero. As in Section 2.6, the value of $\Lambda$ must be zero, and Equation 3.42 simplifies to

$$\left(\frac{\mu_S}{v_S} - \frac{r}{v_B}\right) = \beta_{SM}\sqrt{\frac{v_M}{v_S}}\left(\frac{\mu_M}{v_M} - \frac{r}{v_B}\right)$$ [3.44]

We can rewrite this equation as

$$(\mu_S - R_S) = \beta_{SM}\sqrt{\frac{v_S}{v_M}}(\mu_M - R_M)$$ [3.45]

where the frequency-adjusted riskless rate $R_S$ was previously defined in Equation 3.23.

The coefficient $\beta_{SM}$ on the right hand side of Equation 3.45 is the so-called beta between the calendar-time returns of the stock and the market. It is convenient to define the *frequency-adjusted beta*, $\tilde{\beta}_{SM}$, as

$$\tilde{\beta}_{SM} = \beta_{SM}\sqrt{\frac{v_S}{v_M}}$$ [3.46]

Equation 3.45 can be rewritten as the following intrinsic-time version of the Capital Asset Pricing Model:

$$(\mu_S - R_{SM}) = \tilde{\beta}_{SM}(\mu_M - R_M)$$ [3.47]

It states that speculators who think about risk and return in intrinsic time will conclude that, for a diversifiable stock in a one-factor world, (1) excess return is measured relative to the frequency-adjusted riskless rate, and (2) excess return is proportional to the frequency-adjusted beta times the excess return of the market.

Short-term speculators are often day traders who enter and exit the market for very short periods. For them, the effective frequency-adjusted riskless rate is close to zero, and all profit and loss comes from volatility. In that case, Equation 3.47 simplifies to

$$\mu_S \approx \tilde{\beta}_{SM}\mu_M = \beta_{SM}\sqrt{\frac{v_S}{v_M}}\mu_M$$ [3.48]

Such speculators will expect a short-term return proportional to the traditional beta of the stock, further enhanced by the square root of the ratio of the stock's trading frequency to the trading frequency of the market.

This relation may explain, in a quasi-rational way, why investors jump in to rapidly trading markets and so contribute to the growth of a speculative bubble. They are responding to temperature as though it were risk. An increase in the temperature of one stock in a sector can lead to an increase in the temperature of the entire sector.

# 4 Intrinsic Time, Options Valuation and the Volatility Skew

The original derivation of the Black-Scholes equation was obtained by applying the Capital Asset Pricing Model to both a stock and its option[13]. We can use the same method to derive a simple temperature-sensitive version of the Black-Scholes model.

Consider a stock S and its option C. Applying Equation 3.24 to both of these perfectly correlated securities implies that they share the same intrinsic-time Sharpe ratio, so that

$$\frac{\mu_C - R_C}{\chi_C} = \frac{\mu_S - R_S}{\chi_S} \qquad [4.1]$$

Assume that the option price $C(S, t)$ can be written as a function of the current stock price $S$ and the current time $t$. We can then use stochastic calculus to express $\mu_C$ and $\sigma_C$ in terms of $C(S, t)$, $\mu_S$ and $\sigma_S$ as

$$\mu_C = \frac{1}{C}\frac{\partial C}{\partial t} + \mu_S\frac{S}{C}\frac{\partial C}{\partial S} + \frac{1}{2}\frac{\sigma_S^2 S^2}{C}\frac{\partial^2 C}{\partial S^2} \qquad [4.2]$$

$$\sigma_C = \frac{S}{C}\frac{\partial C}{\partial S}\sigma_S \qquad [4.3]$$

Substitution of these two results into Equation 4.1 leads to

$$\frac{\left(\frac{1}{C}\frac{\partial C}{\partial t} + \mu_S\frac{S}{C}\frac{\partial C}{\partial S} + \frac{1}{2}\frac{\sigma_S^2 S^2}{C}\frac{\partial^2 C}{\partial S^2}\right) - R_C}{\left(\sqrt{v_C}\frac{S}{C}\frac{\partial C}{\partial S}\right)\sigma_S} = \frac{\mu_S - R_S}{\sqrt{v_S}\sigma_S}$$

By simplification one obtains the following modified Black-Scholes equation:

$$\frac{\partial C}{\partial t} + L_S S\frac{\partial C}{\partial S} + \frac{1}{2}\sigma_S^2 S^2\frac{\partial^2 C}{\partial S^2} = R_C C \qquad [4.4]$$

where

$$L_S = \left[R_S\sqrt{\frac{v_C}{v_S}} - \mu_S\left(\sqrt{\frac{v_C}{v_S}} - 1\right)\right] \qquad [4.5]$$

---

13. [Black 1973]

is the effective stock growth rate and $R_C$ is the discount rate in the modified Black-Scholes equation represented by Equation 4.4. Note that the growth rate depends upon $\mu_S$, the expected return for the stock, so that strict risk-neutrality is forsaken.

It is not obvious what value to use for $v_C$, the trading frequency of the option, since options are over-the-counter contracts which can be created at will. One possibility is to take the viewpoint of replication, namely that since options can be created out of stock, it may be reasonable to regard $v_C$ as equal to $v_S$. In that case, $L_S \equiv R_S \equiv R_C = (v_S / v_B) r$. and Equation 4.4 degenerates into a Black-Scholes equation with one effective interest rate $R_S$ which is greater than the true riskless rate $r$ if $v_S$ is greater than $v_B$. If options prices are generated by a Black-Scholes equation whose rate is greater than the true riskless rate, and if these options prices are then used to produce implied volatilities via the Black-Scholes equation with a truly riskless rate, it is not hard to check that the resultant implied volatilities will produce a negative volatility skew.

The intrinsic-time model described here assumes trading frequencies are constant. Perhaps more realistically, the model should be extended to incorporate stochastic trading frequencies. In that case, the calendar-time volatility of the stock, $\sigma_S = \sqrt{v_S} \Sigma_S$, can vary with time and stock price through its dependence on $v_S$, even when the intrinsic-time volatility $\Sigma_S$ remains constant. If trading frequencies depend upon stock prices levels, both increasing as stock prices fall and also varying randomly, then the behavior of calendar-time implied volatilities will incorporate the effects of both stochastic- and local-volatility models. Some of these approaches, which lie outside the scope of this paper, have been recently investigated[14].

---

14. See [Madan 2000] and [Howison 2001].

# 5  Conclusion

In this paper we have derived the consequences of two hypotheses for the relationship between risk and return.

The first hypothesis states that assets with the same risk must have the same expected return. From this we derive the well-known invariance of the Sharpe ratio for uncorrelated stocks, as well as the traditional Capital Asset Pricing Model for stocks correlated with a single market factor.

The second hypothesis is a conjecture, namely, that short-term speculators pay attention to risk and return in intrinsic time. Combining both hypotheses leads to an alternative, more behavioral version of the Sharpe ratio and the Capital Asset Pricing Model.

For uncorrelated stocks, the expected short-term return of a stock is found to be proportional to the temperature of stock, where temperature is the product of the usual stock volatility and the square root of its trading frequency. For stocks correlated with a market factor, the modified Capital Asset Pricing Model replaces the traditional $\beta$ that measures the ratio between a stock's excess return and that of the market by $\tilde{\beta} = \beta(\nu/\nu_M)$, where $\nu$ is the stock's trading frequency and $\nu_M$ is that of the market.

These results, if true, help to explain how the rapid trading of stocks leads investors to imagine that temperature and trading frequency, rather than unalloyed volatility, is relevant to short-run stock returns. One can begin to test these relationships by examining the realized returns of stocks during speculative periods, and examining their correlation with trading frequency.

Finally, we have shown that the theory of intrinsic time can be extended to include options valuation, perhaps accounting for part of the volatility skew.

The Appendix provides a simple illustration of how market bubbles can be caused by investors who, while expecting the returns traditionally associated with observed volatility, instead witness the returns induced by short-term temperature-sensitive speculators.

### Acknowledgements

I am grateful to Sid Browne, Steven Posner, Bill Shen and Leon Tatevossian for helpful comments. I am also glad to thank Parameswaran Gopikrishnan for many useful discussions, and for remarks on the manuscript.

# Appendix: How Bubbles Inflate

A simple model of the interactions between market participants can illustrate how the presence of intrinsic-time-based speculators can affect prices. For simplicity, we assume interest rates are zero.

Consider a stock whose trading frequency is $\nu$ and whose intrinsic-time volatility $\Sigma$ is independent of time and trading frequency. Speculators in this stock will expect the instantaneous return of Equation 3.24, namely

$$\mu_I = \Lambda\sigma\sqrt{\nu} = \Lambda\Sigma\nu \qquad [A1]$$

Suppose that because speculators buy the stock, its price rises to fulfil their expectation, so its realized instantaneous return is indeed $\mu_I$.

Calendar-time-based investors, believing in the more classical risk-return relationship of Equation 2.14, will expect a return

$$\mu_C = \lambda\sigma = \lambda\Sigma\sqrt{\nu}. \qquad [A2]$$

Observing the realized return $\mu_I$, these investors will perceive an excess instantaneous return

$$\Delta = \mu_I - \mu_C = \Lambda\Sigma\nu - \lambda\Sigma\sqrt{\nu}. \qquad [A3]$$

The perception of this excess return can motivate some small proportion of calendar-time investors to buy the stock. Suppose that the number of investors attracted per unit time is proportional to the magnitude of the observed excess return, so that

$$\frac{\partial\nu}{\partial t} = \alpha\Delta = \alpha\Sigma(\Lambda\nu - \lambda\sqrt{\nu}) \qquad [A4]$$

For large calendar time $t$, the first term on the right-hand-side of the above equation becomes dominant, and $\nu$ grows exponentially large, as given by

$$\nu \sim \exp(\alpha\Lambda\Sigma t) \qquad [A5]$$

From Equation A1, $\mu_I = \Lambda\Sigma\nu \sim \Lambda\Sigma\exp(\alpha\Lambda\Sigma t)$, so that the stock's instantaneous return grows exponentially with time. The stock's price therefore inflates through time at the doubly exponentiated rate $\exp(\exp(\alpha\Lambda\Sigma t))$,

This rate is unsustainable. As time increases, the number of investor willing to become speculators saturates, and the number of new trades per second will fail to keep up with the rate demanded by Equation A5.